\documentclass[preprint,12pt]{aastex}

\begin{document}

\newcommand\tna{\,\tablenotemark{a}}
\newcommand\tnb{\,\tablenotemark{b}}
\newcommand\tnc{\,\tablenotemark{c}}
\newcommand\tnd{\,\tablenotemark{d}}
\newcommand\tne{\,\tablenotemark{e}}
\newcommand\tnf{\,\tablenotemark{f}}

\title{Debris Disks around Solar-Type Stars: Observations of the Pleiades with \textit{Spitzer Space Telescope}}
\author{J. M. Sierchio\altaffilmark{1}, G. H. Rieke\altaffilmark{1}, K. Y. L. Su\altaffilmark{1}, P. Plavchan\altaffilmark{2}, J. R. Stauffer\altaffilmark{3}, N. I. Gorlova\altaffilmark{4}}
\altaffiltext{1}{ Steward Observatory, University of Arizona, Tucson, AZ 85721}
\altaffiltext{2}{ Infrared Processing and Analysis Center, California Institute of Technology, MC 100-22, Pasadena, CA 91125}
\altaffiltext{3}{ {\it Spitzer} Science Center, California Institute of Technology, MC 103-33, Pasadena, CA 91125}
\altaffiltext{4}{ Instituut voor Sterrenkunde, K.U.Leuven, Celestijnenlaan 200D, 3001 Leuven, Belgium}
\email{jsierchi@email.arizona.edu}

\begin{abstract}
We present {\it Spitzer} MIPS observations at 24 $\mu$m of 37 solar-type stars in the Pleiades and combine them with previous observations to obtain a sample of 71 stars. We report that 23 stars, or 32\% $\pm$ 6.8\%, have excesses at 24 $\mu$m at least 10\% above their photospheric emission. We compare our results with studies of debris disks in other open clusters and with a study of A stars to show that debris disks around solar-type stars at 115 Myr occur at nearly the same rate as around A-type stars. We analyze the effects of binarity and X-ray activity on the excess flux. Stars with warm excesses tend not to be in equal-mass binary systems, possibly due to clearing of planetesimals by binary companions in similar orbits. We find that the apparent anti-correlations in the incidence of excess and both the rate of stellar rotation and also the level of activity as judged by X-ray emission are statistically weak.
\end{abstract}

\section{Introduction}
Understanding the evolution of planetary systems has long been central to our speculations about our place in the universe and is a challenging and rapidly advancing topic in astronomy and planetary science. Several hundred extrasolar planets have been discovered in the last 15 years, establishing that planetary systems form frequently. We have also significantly expanded our understanding of the planet formation process, through measurements of protoplanetary disks in the critical mid-infrared through submillimeter regimes \citep[e.g.,][]{andrews2005,silverstone2006,hernandez2006} and development of detailed two- and three-dimensional numerical models to simulate and interpret these results \citep[e.g.,][]{alexander2008,dullemond2007}. However, after protoplanetary disks dissipate at about 5 million years, the processes leading to planet formation and its remaining signatures become almost invisible to most types of observation.

Planetary debris disks, resulting from collisional activity among planetesimals \citep{wyatt2008}, are a notable exception and thus are our best current means to characterize planet system evolution. Debris disks are readily detected through the infrared emission from dust particles, which are produced in collisional cascades initiated among larger bodies, e.g., through gravitational stirring by planets. The generation of dust can be further enhanced when the particles become small enough that non-gravitational forces become significant, leading in extreme cases to avalanches of particle generation \citep{grigorieva2007}. The dust is cleared from the disk on time scales of a thousand to a million years, so it must be replenished. Thus, debris disks are an effective indirect means to probe the current level of collisions in planetary systems.

One application of debris disk studies is to characterize the collisional activity as planet systems age \citep[e.g.,][]{habing2001,spangler2001}. Previous work has shown that debris disks tend to decay with time \citep{rieke2005,siegler2007}. The excellent sensitivity of the Multiband Imaging Photometer on {\it Spitzer} \citep[MIPS;][]{rieke2004} and, more importantly, the accurate photometry it can deliver, have expanded such studies substantially \citep[e.g.,][]{rieke2005,su2006,gorlova2006,siegler2007,meyer2008,trilling2008,hillenbrand2008,carpenter2009a,balog2009}. In addition to accurate infrared photometry, such studies depend on the accurate determination of stellar ages. Substantial uncertainties remain in even the best age determinations for field stars \citep[e.g.,][]{mamajek2008}. Therefore, measurements of excesses in stellar clusters, where ages are better determined, can refine the estimates of decay trends. However, the number of clusters close enough for the measurements to probe to solar masses and below is limited, and characterizing the decay trend and other disk parameters is therefore limited by poor statistics \citep{gaspar2009}. For these reasons, it is important to observe thoroughly the small number of nearby rich clusters available. Only through such measurements can we test whether there are differences in disk behavior around different stellar types, with their accompanying differences in luminosities, masses, and presumably protoplanetary disk masses. For example, initial studies failed to find significant differences in infrared excess decay with respect to spectral type \citep[e.g.,][]{gorlova2006,siegler2007}, but with improved statistics they seem to be emerging \citep{gaspar2009}. Removing age as a parameter also allows us to study other parameters affecting debris disk evolution, such as the various mechanisms for grain removal.

This paper completes our analysis of {\it Spitzer} debris disk surveys in the Pleiades. We report 24 $\mu$m measurements of a sample of 37 stars of late F to early K type in this cluster. We combine these measurements with previous surveys by \citet{stauffer2005} and \citet{gorlova2006} to assemble a combined sample of 71 such stars. Our final excess rate provides the highest-weight determination available for debris disk behavior at 115 $\pm$ 10 Myr, the age of the Pleiades \citep{meynet1993,stauffer1998,martin2001}. In Section 2, we describe our observations and sample selection. In Section 3, we describe how we obtained excess ratios for our sample using both color-color plots and Kurucz model-fitting. In Section 4, we interpret these excesses through comparisons with similar studies of Praesepe, Blanco 1, and NGC 2547. We also present an analysis of binarity, rotation, and stellar wind drag, as possible ways for the dust to be removed. Finally, in Section 5 we conclude with a summary and some possibilities for future work.

\section{Observations, Data Reduction, and Sample Selection}
Table 1 lists our sample stars, with notations from \citet[HII:][]{hertzprung1947}, \citet[AK:][]{artyukhina1968} and \citet[][\citealt{vanleeuwen1986}]{pels1975}. Their Pleiades membership is based on criteria similar to those used by \citet{stauffer2005}; i.e., they are likely cluster members based upon radial velocities, proper motions, chromospheric and coronal activity indicators, and lithium abundance \citep{stauffer1987,soderblom1993,queloz1998,rosvick1992}. These stars are mostly of spectral types F5 to K1, when types are available; we also required 1.05 $< V-K_S <$  2.15, appropriate for this range of types \citep{tokunaga2000}. 

A few stars need to be discussed individually. HII 3031 is of indicated type F2 and fails the color selection. AK1B590 is indicated to be type F2 but passes the color test, so we have retained it. HII 1139 and Pels 128 were also rejected because the photometry may be influenced by close stars of similar brightness and HII 2172 was rejected because of a structured background. Thus, we report the photometry of HII 3031, HII 1139, HII 2172, Pels 128, and Pels 173\footnotemark, but do not include them in the analysis of the final sample. The remaining 32 objects are termed the new sample.

\footnotetext{Pels 173 was eliminated because of general confusion about its nature. It is generally considered to be the C component of the triple system, BD +22 617 = HD 25201. The A and B components are a close B9/A binary with proper motion and photometric characteristics consistent with membership in the Pleiades, but with a parallax that puts them significantly behind the cluster. The C component, about 1' removed, is of spectral type F2 and has photometric colors consistent with this designation, but is too faint by $\sim$ 0.6 mag for an F2V star within the Pleiades. The most likely designation for these three stars is that they in fact are a physical system but are behind the Pleiades, and therefore, for our purposes, are not useful because there is no good age estimate. Many of the discussions of this star in the literature mix properties of the different components in non-physical combinations.}

We used the MIPS in photometric mode to measure the 24 $\mu$m emission from the stars in the new sample (PID: 30503) with final integration times of 93--313 s depending on the source brightness. The data were reduced with the MIPS Instrument Team Data Analysis Tool \citep[DAT;][]{gordon2005} with the specific prescription for maximum photometric accuracy described by \citet{engelbracht2007}. The calibration is based on \citet{rieke2008}. We extracted photometry using point-spread function (PSF) fitting with a smoothed theoretical PSF generated by the STinyTim program \citep{krist2006}. We also compared the results for consistency with aperture photometry of the same star and examined each field for nearby objects or background structures that might compromise the photometry, leading to rejection of HII 1139, 2172, 3031, and Pels 128 mentioned above. Most of the stars reside far from the cluster center, reducing the risk of poorly measured or confused 24 $\mu$m emission. The sample properties are summarized in Table 1.

We also include in our analysis other Pleiades members measured by \citet{stauffer2005} and \citet{gorlova2006}. The full sample of \citet{gorlova2006} includes Pleiades members of B-type and later. We took from it all stars where the appropriate intrinsic colors matched the color selection in our new sample. Where spectral types are available, they generally agree with this color selection, although there are two stars within the color selection (HII 1338 and 1912) that are indicated to be F3 and F4, respectively.  Although their relatively red colors for their types might arise from reddening, there might also be type errors, so we retained them in the sample. We rejected stars where the indicated noise exceeded 4\% of the signal (HII 1095, 1207, 1794, 2311, 2341), since our goal is to detect excesses at this level ($1\sigma$). All together, the combined sample for analysis has 71 members, consisting of the 32 in our new sample observed for this paper plus all retained stars from \citet{gorlova2006} and \citet{stauffer2005}. Table 2 lists this complete sample with flux measurements.

\section{Determination of Excesses}
Identifying 24 $\mu$m excesses requires an accurate extrapolation of the photospheric output to this wavelength. We define the ratio of the observed 24 $\mu$m flux density to the extrapolated photospheric flux density, the excess ratio, as an indicator of the level of excess emission (no excess and no measurement errors would yield an excess ratio of unity). We first describe how we carried out this step for the new sample. We then discuss the steps to put the studies of \citet{stauffer2005} and \citet{gorlova2006} on the same scale to provide a uniform combined sample.

\subsection{The New Sample}
We assembled a database including available optical and near-infrared photometry for the stars in this sample. We used these data for two independent approaches to extrapolate the photospheric spectral energy distributions (SEDs), one based on fitting stellar SED models and the other based on applying photometric colors. Comparing the results of these methods shows that they give similar performance, so our final excess ratios are based on the average of the two SED extrapolations.

\subsubsection{SED Fitting}
The database listing for each star included, if found, spectral type, extinction, temperature, and the following photometry: Johnson $UBVRI$, Two Micron All-Sky Survey (2MASS) $JHK_S$, Hipparcos $BV$, and Str\"{o}mgen $uvby$. Almost all the stars had $BVRJHK_S$ photometry available, most had $U$ and $I$, and a few had the other bands. Table 1 presents our database of photometry, effective temperatures, surface gravities, spectral types, and metallicities used for SED fitting.  

The SED fitting was based on Kurucz models \citep{castelli2004} and utilized an interactive code written in IDL. To create the SEDs, we used the constraint that the Kurucz model-fitting program fit the available photometry from 0.5 to 5 $\mu$m. We recorded the Hipparcos $B$ and $V$ photometry separately and included it in the fits as independent points from the literature $BV$ photometry, unless the literature photometry was based on those measurements. The effective temperatures listed in Table 1 were used as initial estimates in the SED fitting procedure. For each star we searched for models at $\pm$ 750 K centered near the expected effective temperature to obtain the best-fit SED.

\subsubsection{Photometric colors}
Extrapolation of the photospheric SED using photometric colors, e.g $V-K_S$ versus $K_S$-[24], is a relatively easy method to apply to large samples of stars. It has the convenient feature that it is self-calibrating. Since many stars can be anticipated to have no excess emission at 24 $\mu$m, there is a strong peak at small values of the excess ratio that can be set to one. There is a risk that small excesses will be lost through this approach. In comparison, SED fitting as in Section 3.1.1 has the advantage that it retains the absolute SED level, but with the disadvantage that the level deduced depends on the accuracy of the theoretical photosphere models. In this paper, we will place color-based extrapolation on an absolute scale and then compare the results with those from SED fitting to probe these advantages and disadvantages.

We used the FEPS sample \citep{carpenter2008} to probe different approaches to using photometric colors, after checking that the 24 $\mu$m photometry in that paper is consistent with ours. For this sample, we can use 2MASS $JHK_S$ photometry to anchor photospheric predictions, since almost all the members are fainter than the 2MASS saturation limit. The FEPS team also obtained IRAC photometry at three bands for the sample. To select a subsample that should have very few stars with excess emission, we used chromospheric activity as an indicator of stellar age and accepted only those stars indicated to be older than 500 Myr in the calibration of \citet{mamajek2008}. We used the width of the peak in excess ratio, $W_{ex}$, as an indicator of the quality of the extrapolation; this metric is not influenced by a small number of outliers (e.g., stars with excess emission).

\citet{carpenter2008} derive a number of small corrections to the IRAC and MIPS photometry based on the behavior of the sample as a whole. We found that application of these corrections either in the IRAC bands or at 24 $\mu$m increased $W_{ex}$. The best performance was obtained by using just the uncorrected 8 $\mu$m photometry, yielding an rms scatter of 1.6\% in the extrapolated flux density at 24 $\mu$m compared with the measured values.

For many stars listed in Table 2, including the new sample of Pleiades stars, IRAC photometry is not available; 2MASS (or other $JHK$) photometry must be used for SED extrapolations. We found that this extrapolation was most accurate when the $K_S$-band measurement was combined with the $H$-band measurement using the nominal $H-K$ color for the spectral type of the star and weighting by the quoted 2MASS errors. To do this, we plotted $V - K_S$ versus $H-K_S$ and used a fit to determine $V-K_S$, which we used to estimate the stellar types. Addition of the $J$-band data decreased the accuracy, presumably because the $J-K_S$ color is a relatively strong function of spectral type and the derived spectral types have significant errors. The result using just the $H$- and $K_S$-band data provided rms deviations of 2.5\% in the extrapolated flux density at 24 $\mu$m, i.e., about 1.5 times the scatter obtained with the 8 $\mu$m band. This study allowed us to set the value of 89 for the ratio of $K_S$-band to 24 $\mu$m flux densities that gives an excess ratio of one for stars with no excess. This value agrees to within 1.4\% with the similar value from the calibration of \citet{rieke2008}, that is, within the quoted error there of 2\%. We use the value derived from the FEPS data to analyze our observations of the new sample.

\subsubsection{Final Excess Ratios}
For the new sample, the excess ratios determined by SED fitting and by photometric colors are of comparable quality. They are generally in agreement with an rms difference of 3\%. We therefore averaged the two determinations. The final value of the rms scatter for the stars with no evidence for excesses is 3.6\%. Given typical errors for the 2MASS $K_S$ magnitudes (even after combining with the $H$-band magnitudes of 2\% and the expected noise level of the observations $\sim$ 3\%), this value is reasonable. It is only slightly worse than the noise level achieved on much brighter field stars \citep[e.g.,][]{su2006,trilling2008}, showing that no additional error sources intrude at the much lower signal levels of the Pleiades stars.

\citet{stauffer2005} used IRAC 8 $\mu$m photometry for their SED extrapolations. We put their results on the same scale as those for the new sample by applying the corrections derived in \citet{rieke2008}. We augmented the results of \citet{gorlova2006} with 2MASS $H$-band photometry converted to the $K_S$-band as described above, in all cases where the indicated $K_S$-band error was greater than 0.02. Table 2 includes the 24 $\mu$m fluxes and excess ratios for the full sample.

The distribution of excess ratios for the combined sample is plotted in Figure 1, which also shows a Gaussian fit to the distribution. The fit was optimized by computing chi-squared over a window centered at the derived fit center and extending to the point at $\sim$ 15\% of the distribution peak on either side. This approach was designed to reduce the effect on the fitted parameters by stars with excesses. The final fit is centered at 0.998 with a standard deviation of 0.032.

\section{Results}
From Figure 1, excess ratios $>$ 1.1 are identified as having infrared excesses at 24 $\mu$m. This corresponds to 3$\sigma$ of the Gaussian fit shown. The incidence of excesses for ratios $>$ 1.06 can be estimated by subtracting the fit to the no-excess distribution from the observed numbers of stars with excesses in this range. Out of a total combined sample membership of 71 stars, 23 have excess ratios exceeding the 3$\sigma$ level. Fourteen have ratios $>$ 1.16 (or $>$ 5$\sigma$) for comparison with the study of Praesepe by \citet{gaspar2009}. An additional six stars may have excess ratios between 1.06 and 1.10. By definition, subtracting these numbers from the observed distribution leaves the values fitted adequately by the Gaussian for no excess. Since its center is not significantly offset from one, it is likely that these remaining 42 stars have very little excess emission down to a low level. Expressed as percentages, 20\% $\pm$ 5.3\% have excess ratios $>$ 5$\sigma$, 32.4\% $\pm$ 6.8\% have them $>$ 3$\sigma$, and about 41\% $\pm$ 7.6\% have them $>$ 1.06 ($\sim$ 1.7$\sigma$). These values are higher then previous estimates for the incidence of excess in the Pleiades, but consistent with the old ones within the errors and given the identification of excess at smaller flux levels in this analysis. The remaining 59\% have no excess exceeding the 1.7$\sigma$ level.

Because we include in our analysis stars from previous studies, we will address some possible biases in our results. We include HII 1338 and 1912, which do not meet the color-selection criteria. Since early-type stars in general may have a higher incidence rate of 24 $\mu$m excesses than later type ones, a bias toward stars with excess emission might result. However, neither of these stars has an excess, adding no bias to the sample in that regard. We also rejected other stars due to low signal-to-noise ratio (see Section 2). Rejecting stars with low signal-to-noise ratio might bias the sample in favor of excesses; however, only one of these stars (HII 1095) has any indication of an excess (in this case at the $4\sigma$-level of significance), so the incidence of excesses among these stars is approximately the same as we derive for the full sample and no significant bias is likely. Therefore, we are confident that the excess rates derived above are correct for our sample.

\section{Discussion}
Our study of the Pleiades probes the debris disk characteristics in a large sample of solar-type stars of well-determined age. We compare our sample to similar studies done with A stars and the Blanco 1 open cluster. We show that debris disks emitting at 24 $\mu$m are still common at 100 Myr around both A- and solar-type stars, which provide an important constraint on theoretical models for their dissipation and replenishment. We also use this sample to examine possible correlations with other stellar parameters. Normally, the dramatic decay of infrared excesses with age, combined with the uncertainties in stellar ages, can mask important second-parameter effects. However, since age variation is removed in the studies of the Pleiades and other open clusters, they provide the best opportunity to examine other influences on debris disk behavior.

\subsection{Comparison of Incidence of Excesses}
Our value for the incidence of excesses in the Pleiades can be compared with that of \citet{stauffer2009} on the Blanco 1 cluster. This cluster is similar in age to the Pleiades ($\sim$ 100 Myr) but about twice as far away ($\sim$ 250 pc). Within the spectral type range for our Pleiades study (1.05 $\le V-K_S \le$ 2.15), there are 18 Blanco 1 members observed at 24 $\mu$m. Two of them, W91 and W99, are identified as having 24 $\mu$m excesses at $>$ 3$ \sigma$ \citep{stauffer2009}. Three more, W38, W53, and Z5102, have excesses indicated at $>$ 2.5 $\sigma$. The method used to estimate $\sigma$ in this paper, comparing photometry of sources observed in more than one pointing, is very conservative as applied to sources targeted for photometry. The latter objects are centered in the field and measured in exactly the way used for calibration stars, on which the standard sensitivity estimates are based. The former objects tend to lie near field edges and would be expected to have lower-accuracy measurements. In fact, the \textit{Spitzer} Science Center sensitivity estimator suggests that the 1-$\sigma$ level for the targeted objects in the \citet{stauffer2009} survey should be about 25 $\mu$Jy ([24] = 10), which is nearly twice as accurate as indicated in \citet{stauffer2009}.  We therefore consider W38, W53, and Z5102 to have reliably measured excesses also. We adopt a net count of 5 excesses among 18 stars, or an incidence of $\sim$ 28\%. This result is close to the one for the Pleiades, albeit at much lower statistical significance.

We can also compare our Pleiades sample with the members of the sample of A stars selected from \citet{su2006} whose ages are between 50 and 300 Myr. This study has a \textquotedblleft zero-point\textquotedblright excess ratio for no excess of 0.981 and 1$\sigma = 0.026$, compared with 0.998 and 0.032, respectively, for our Pleiades sample. Therefore, to compare at similar detection thresholds, we determine the number of A-stars with excess ratios above 1.08 and 1.13 for 3$\sigma$ and 5$\sigma$, respectively. Of 69 A stars within this age range, 21 have excess $>$ 5$\sigma$ and 25 have them $>$ 3$\sigma$, corresponding to 30\% $\pm$ 6.6\% and 36\% $\pm$ 7.2\%, respectively. The latter value is slightly higher than the corresponding one for solar-type stars, i.e. (combining the Pleiades and Blanco 1), 28 of 89, or 31\% $\pm$ 5.9\%, but the difference is within the statistical uncertainties. 

The relative weakness of this dependence on spectral type at the age of the Pleiades can be contrasted with the virtual complete absence of 24 $\mu$m excesses around solar-type stars at 750 Myr \citep{gaspar2009}. It appears that old A-type stars have a higher incidence of excesses, although they too experience a rapid decline in this regard after $\sim$ 100 Myr \citep{gaspar2009}. There may be a relatively simple explanation for this type of difference in excess evolution. The A-type stars have typically 2.5 times the mass and 25 times the luminosity of the solar-type ones. The thermal equilibrium distance from the star for grains emitting at 24 $\mu$m will be five times greater for the A stars, and the Keplerian orbital velocities of the grains will be square root of two less. Thus, the collisional evolution time scales for the portions of the debris systems that dominate the 24 $\mu$m emission should be longer for the A stars than for the solar-type ones. These scalings also show that all the dust production and clearing time scales will be longer for the A-type stars \citep{dominik2003}, again consistent with the slower decay of the excesses around them. We can also relate this to our own Sun. Combining our excess rate at 100 Myr with that at 750 Myr, we estimate that only a small fraction of solar-type stars at the age of the Sun, if any, will have detectable excesses at 24 $\mu$m, assuming a steady decay rate. An analysis of the excess decay time scales as a detailed function of stellar mass, e.g., using the full {\it Spitzer} data set, may be able to test this basic hypothesis and to refine it into a test of models for debris disk evolution.

\citet{carpenter2009b} present a similar study, comparing the excess fractions of early- and solar-type stars at various ages by compiling data from several previous works (see Figure 6 and Table 4 in their paper). From Figure 6 in their paper, at 100 Myr, the excess fractions between early- and solar-type stars differ significantly, with the latter type having the lower excess fraction. However, \citeauthor{carpenter2009b} define solar-type stars as being from G0 to K5 in spectral type, whereas we have defined solar type as F5 to K1, placing a G2 star more in the center of the range, in regard to mass. Using color as an approximate indication of spectral type, we count 27 stars in our full sample as F-type, and the remaining 44 as G or K type. Of the 27 F-type stars, 12 have excesses giving a fraction of 44\%. Of the G- and K-type stars, 11 have excesses giving a fraction of 25\%. We attribute the difference in results between our paper and \citet{carpenter2009b} as due to the difference in definition of solar-type stars.

\subsection{Excesses and Binarity}
\citet{stauffer2009} found that binary stars in Blanco 1 and NGC 2547 tended not to have 24 $\mu$m excesses. \citet{gorlova2006} previously found similar results for stars in the Pleiades. We use similar methods to probe this correlation for our combined sample. Figure 2 shows $V$ versus $B-V$ for the stars listed in Table 2. Possible binary systems were found fitting the photometry to the single star locus found in \citet{stauffer2007}. The difference between the observed $V$ magnitude and the single-star locus, or $\Delta V$, for each star is listed in Table 2; stars with $\Delta V$ of greater than 0.2 are likely to be binaries. To test whether binarity affects the excess rate for our sample, in Figure 3 we show excess ratio versus $\Delta V$. The trend is very similar to that reported by \citet{stauffer2009} and confirms the anti-correlation between binarity and 24 $\mu$m excess. Both stars with extreme excesses, AK II 437 and HII 1132, follow this trend, although they are not shown in Figure 3. To test this trend for the higher-mass binary systems, we used a Kolmogorov Smirnov (KS) test, which revealed an 8\% probability that there is no anti-correlation between excess and high-mass binarity, assuming a cutoff of 0.4 for $\Delta$V. 

Most of the Pels and AK stars have not been well-studied for binarity. However, separations for stars previously identified as binaries are listed in Table 2. The typical separation is less than 0.5 arcsec, which corresponds to $\sim$ 70 AU at the distance of the Pleiades. Given the separations, it is plausible that the binary companions are disrupting any possible disks around these stars. In fact, Figure 4 of \citet{trilling2007} shows a general absence of 24 $\mu$m emission for field binaries with similar separations to those that are typical in the Pleiades, and they explain the effect as being a result of the disruptive effect of a binary companion at a distance from the star where a debris disk could typically be located.

\subsection{Correlations of Excesses with X-ray Activity and Rotation}
In general, for solar-type stars, rotation rate correlates with X-ray luminosity \citep{noyes1984}; both correlate inversely with the age and may correlate directly with the strength of the stellar wind \citep{wood2005,holzwarth2007,cranmer2008}. There are a variety of reasons to expect correlations among these parameters and the incidence of infrared excesses. For example, \citet{currie2008} find an inverse correlation between the stellar rotation rate and the amount of material available to produce a debris disk among early-type stars in the young cluster NGC 2232, and suggest that massive protostellar disks allow stars to spin down efficiently so that an inverse correlation is to be expected. A possible issue with applying this argument to excesses at 24 $\mu$m is that most of the mass in a debris disk is typically in the cold component that dominates the 70 $\mu$m emission, so the correlation may not be apparent at the shorter wavelength. Another possibility is that strong stellar winds remove debris dust quickly \citep{plavchan2005, minato2006}. The difficulty in confirming this prediction lies in the uncertainties in determining stellar winds, as is made clear in the review by \citet{cranmer2008}. The number of stars with direct wind measurements is small; they indicate a roughly linear dependence on X-ray emission but with a substantial deviation downward at high X-ray luminosity \citep{wood2005}.  

By removing the age dependence of the infrared excesses, observations in clusters let us probe for any relations with these other parameters. Below, we consider rotation and X-ray surface brightness (and presumably stellar winds). 

\subsubsection{Rotation}
We use the measured $v$sin$i$ as an indicator for stellar rotation rate. In Figure 4, we show the excess ratio at 24 $\mu$m versus $v$sin$i$ for the stars in Table 2. Each star with a significant excess has $v$sin$i$ $<$ 50 $km s^{-1}$, including both stars with extreme excesses (AK II 437 and HII 1132). \citet{stauffer2009} found a similar effect in the Blanco 1 data. \citet{rebull2008} also found a slightly inverse trend between excess and rotation for G, K, and M stars in the $\beta$ Pic moving group, but suggested as an alternative explanation that it could be due to stellar wind. 

However, in all these cases, the number of stars in the samples also decreases with increasing $v$sin$i$, making it necessary to test for selection bias.  To do so, we have combined the Pleiades and Blanco 1 samples (after removing from the former stars with extreme excesses that probably arise through recent large events (HII 1132 and AKII437)). A K-S test gives a 38\% probability that the correlation arises by chance, using 50 $km s^{-1}$ as the boundary between slow and fast rotation. After the eliminations listed above, there are only 7 stars with $v$sin$i$ above that boundary, compared to 72 stars below. With a cutoff of 30 $kms^{-1}$, the probability increases to 49\%. Figure 4 is suggestive of a correlation between 24 $\mu$m excess and rotation because none of the fast rotators have excesses; however, because there are so few stars with large $v$sin$i$, the apparent correlation could be the result of small number statistics. 

Stellar rotation has also been studied in a number of young clusters through high accuracy photometry \citep{aigrain2007}. However, the overlap between the stars with measured rotation periods and those studied for excesses with {\it Spitzer} is inadequate to extend this test. Within the limitations of current data, we conclude that there is no evidence for an inverse correlation of infrared excess at 24 $\mu$m with rotation, at least for solar-type stars of $\sim$ 100 Myr age.

\subsubsection{Stellar Winds}
The Pleiades stars generally show an inverse relationship between X-ray luminosity and the presence of 24 $\mu$m excesses. Figure 5 shows the excess ratio at 24 $\mu$m versus X-ray surface brightness for members of the Pleiades, NGC 2547, and Blanco 1, all of which seem to show this effect. Similar behavior was found previously for much younger stars in the Scorpius-Centaurus OB association \citep{chen2005}.

Such behavior might be associated with stellar winds. There are several mechanisms for clearing dust grains in debris disks, including particle-particle collisions and, when the grains are small enough, Poynting-Robertson drag, stellar wind drag, and photon pressure \citep[][and references therein]{plavchan2009, gustafson1994}. The relative roles of these grain-removal mechanisms depend on the age and spectral type of the star \citep[][2005]{plavchan2009}. \citet{dominik2003} and \citet{wyatt2005} show that grain-grain collisions are more important than Poynting-Robertson drag for grain removal in typical mature debris disks. However, winds act in a similar way to Poynting-Robertson drag to cause grains to spiral into the star. The strong winds typical in young stars can become the dominant grain removal mechanism for winds greater than three times the solar value, so long as the amount of dust in the relevant part of the debris system is moderate \citep{minato2006,plavchan2009}. \citet{plavchan2009} show that stellar winds affect the evolution of debris disks around K and M dwarfs, and suggest that they could also be important for young solar-type stars.

We have investigated this possibility by computing approximate X-ray surface brightnesses, $F_x$  (which should be roughly correlated with wind strength), and comparing these values with excess ratio. We obtained stellar surface areas ($A_*$) by estimating stellar luminosities, masses, and temperatures from an isochrone \citep{marigo2008,bonatto2004} fitted to the stars' $M_K$ and $V-K_S$ measurements. Fitting parameters for the isochrones for each cluster are listed in Table 3. In keeping with our sample selection for the Pleiades, we selected from the NGC 2547 and Blanco 1 subsamples stars with 1.05 $< V-K_S <$ 2.15 to limit our analysis to solar-type stars.

We computed the fractional infrared luminosities as $\sim \nu L_\nu(24 \mu m)$, assuming the disk temperatures are about 150 K. This temperature yields a lower limit to the true fractional luminosity. This assumes that the stars are single stars, but we have already shown that another removal mechanism tends to dominate for binaries. Upper limits on the fractional luminosities for those stars with excess ratios less than 1.0 were assumed to have a maximum excess ratio of 1.064, corresponding to 2$\sigma$ above the photosphere. We assumed cluster distances of 407 pc and 250 pc for NGC 2547 and Blanco 1, respectively \citep{mayne2008,panagi1997}. For the Pleiades, we assumed the more traditional distance of 130 pc.

We used the X-ray surface brightnesses to estimate the stellar wind strength. There are indications that any relation between wind and X-ray properties may saturate at large values \citep{wood2005,cranmer2008}, producing significant differences between the observed and predicted wind strengths. But because the saturation level is not well-determined, we use the formula proposed by \citet{wood2005} scaled to 36 Oph to estimate the dependence of the stellar wind strength on the X-ray surface brightness. Below, we discuss each cluster individually.

{\it Pleiades.} We used {\it ROSAT} X-ray fluxes. For HII 1797, we fitted the isochrone to its $B-V$ color instead of $V-K_S$, as 2MASS $K_S$ was not available. Pleiades X-ray luminosities and fractional infrared luminosities are listed in Table 2. Only one star, HII 152, has a significant excess and a strong estimated stellar wind. Neither star with an extreme excess, AK II 437 and HII 1132, has {\it ROSAT} data.

{\it NGC 2547.} For those NGC 2547 members with 24 $\mu$m detections, X-ray luminosities and V magnitudes were obtained from \citet{jeffries2006}. \citet{gorlova2007} identified stars with excesses on a V-K versus K-[24] plot. We found an approximate excess ratio for each star and set 1.10 as the threshold for having a meaningful 24 $\mu$m excess for NGC 2547, similar to that for the Pleiades. The excess ratios and fractional luminosities for NGC 2547 members are listed in Table 4 by right ascension (R.A.) and declination (decl.). Two stars in NGC 2547 (R.A., decl.: 122.446875, -49.21808333 and 122.5642083, -49.09686111) show strong winds and excesses. There are eight stars in total that meet our V-K color criteria and are reasonably likely to have 24 $\mu$m excesses \citep{gorlova2007}. Assuming upper limits of log$L_X$ $<$ 29.3 for the ones not detected, then six stars with infrared excess have weak winds. This is a conservative estimate of the upper limit on log$L_X$ given that nearly half of the stars in the sample were detected having a log$L_X < 29.3$, and it accounts roughly for the drop-off in sensitivity with off-axis angle of the telescope. 

{\it Blanco 1.} Photometry was obtained from \citet{stauffer2009}. Fractional luminosities were computed using the same method as for the Pleiades and NGC 2547 and are listed by star in Table 5. None of the 19 stars show an excess and a strong indicated stellar wind.

Including stars with upper limits, for all three samples it is uncommon for a star with a strong computed wind to show a significant infrared excess. Since only a modest wind is sufficient to dominate the grain removal \citep{minato2006,plavchan2009}, this result should be independent of the saturation effects. When we use the KS test, the lowest probability that this correlation arises by chance is about 12\%, using $F_X \sim 7.2 \times 10^{6} ergs s^{-1} cm^{-2}$ as the threshold. As a check, we repeated this procedure using our upper limits on the fractional luminosities in place of the excess ratios at 24 $\mu$m and found no evidence for a correlation. Allowing for the extra free parameter associated with adjusting the K-S test to minimize the probability of a chance correlation, these results again suggest that there is no significant relation. 

We also tried to estimate the level of saturation for the wind strengths, assuming that collisional timescales do not differ much from drag timescales (\citep[e.g., see Equation (A23) in][]{plavchan2009}. A significant number of stars have substantial excesses, even though the indicated wind strength would produce subtantial drag. It is therefore likely that the wind strengths are overestimated. However, we cannot say anything more quantitative due to the issues mentioned above as well as errors in our estimates of the fractional luminosities. More observations and/or an alternative estimate of wind strengths are needed to test the influence of stellar winds on infrared excesses.

\section{Conclusions}
We report the results of a new survey of solar-type stars in the Pleiades for 24 $\mu$m excess indicative of circumstellar debris disks. We combine this work with previous surveys \citep{stauffer2005,gorlova2006} to build a sample of 71 solar-type stars in this cluster with sufficiently accurate data to identify excesses as small as 10\% at 24 $\mu$m. Twenty-three of these stars have excesses at this level or above; on statistical grounds, it is likely that about six additional members have excesses in the 6\% to 10\% range, and the remaining 42 stars must have little or no 24 micron excess. The incidence of excesses at 24$\mu$m and at the age of the Pleiades is high, $\sim$ 31\% $\pm$ 6\%. 

We find that the incidence of 24 $\mu$m excesses for solar-type stars in the Pleiades is slightly smaller than for A stars \citep[from a general sample mostly in the field:][]{su2006}. It appears that by an age of $\sim$ 750 Myr, the excesses around solar-type stars have decayed faster than they decay around A stars \citep{gaspar2009}. The effect probably arises through a mechanism that operates relatively slowly, such as systematic velocity differences in 24$\mu$m-emitting zones of the debris disks around the two stellar types and the resulting difference in the speed of debris disk evolution.

Our study of the Pleiades, plus similar work on other clusters, lets us test aspects of debris disk behavior independently of the evolution of these systems with age. We confirm the results of \citet{stauffer2009} that close, high-mass binary systems tend not to harbor debris disks. This behavior is probably associated with binary companions that orbit close to the zone where debris disks tend to lie \citep{trilling2007}. There appear to be anticorrelations between infrared excesses and both rotation ($v$sin$i$) and X-ray luminosity, as also indicated by some previous works. However, we find that these results are not statistically significant and may arise instead from selection effects within the debris disk sample. The excesses around stars with indicated strong winds (from X-ray surface brightnesses) suggest that the wind strengths may be overestimated.

\acknowledgements
The authors thank Andr\'{a}s G\'{a}sp\'{a}r for helpful discussions and the anonymous referee for comments which improved the paper. This research has made use of the SIMBAD database and the VizieR catalog access tool, operated at CDS, Strasbourg, France. This work also makes use of data products from the Two Micron All Sky Survey, which is a joint project of the University of Massachusetts and the Infrared Processing and Analysis Center/California Institute of Technology, funded by the National Aeronautics and Space Administration and the National Science Foundation. This work is also based, in part, on observations made with {\it Spitzer Space Telescope}, which is operated by the Jet Propulsion Laboratory, California Institute of Technology under NASA contract 1407. This work supported by contract 1255094 from Caltech/JPL to the University of Arizona.

\begin{deluxetable}{llcccccccccccccl}
\rotate
\tabletypesize{\scriptsize}
\tablecolumns{16} 
\tablewidth{0pt}
\tablecaption{Basic Properties of Pleiades Stars: New Sample}
\tablehead{\colhead{Target}                                         &
           \colhead{Other Name}                                     &
	   \colhead{S. T.\,\tablenotemark{a}}                       &
	   \colhead{$T_{eff}$}                                      &
	   \colhead{Log g\,\tablenotemark{b}}                       &
	   \colhead{[Fe/H]\,\tablenotemark{c}}                      &
	   \colhead{$U$}                                            &
	   \colhead{$B$}                                            &
	   \colhead{$V$}                                            &
	   \colhead{$R$}                                            &
	   \colhead{$I$}                                            &
	   \colhead{$J$}                                            &
	   \colhead{$H$}                                            &
	   \colhead{$K_S$}                                          &
	   \colhead{AOR Key}                                        &
	   \colhead{Ref.}                                           }
\startdata
AK\,IA\,36 & HIP\,17317 & - & 5500\tnd & - & - & - & 11.084 & 10.384 & 10.00 & 9.65\tne & 9.117 & 8.868 & 8.758 & 18304256 & -,-,-,-,-,1,1,2,3,4,4,4 \\
AK\,1A\,56 & BD+21 508 & F8 & 6200 & - & - & - & 11.119 & 10.474 & 10.33 & - & 9.086 & 8.819 & 8.777 & 18304512 & 5,6,-,-,-,7,7,8,-,4,4,4 \\
AK\,1A\,76 & HD\,23312 & F5 & 6310 & 4.44 & - & 10.00 & 9.98 & 9.49 & 9.05 & 8.74 & 8.52 & 8.349 & 8.278 & 18306560 & 5,9,9,-,10,10,10,10,10,4,4,4 \\
AK\,1A\,317 & HD\,24463 & G0 & 6030 & - & - & - & 10.234 & 9.718 & 9.28 & - & 8.687 & 8.48 & 8.424 & 18303744 & 5,6,-,-,-,1,1,2,-,4,4,4 \\
AK\,1B\,7 & HD\,22627 & G0 & 6000\tnd & - & - & - & 10.417 & 9.831 & 9.47 & - & 8.728 & 8.54 & 8.434 & 18303488 & 11,-,-,-,-,1,1,2,-,4,4,4 \\
AK\,IB\,8 & HIP\,17044 & - & 5700\tnd & - & - & - & 11.009 & 10.391 & 9.93 & 9.72\tne & 9.232 & 8.999 & 8.905 & 18306048 & -,-,-,-,-,1,1,8,3,4,4,4 \\
AK\,1B\,365 & HD\,23598 & F8 & 6030 & - & - & - & 10.393 & 9.822 & 9.38 & - & 8.72 & 8.463 & 8.435 & 18304000 & 11,6,-,-,-,12,12,2,-,4,4,4 \\
AK\,1B\,590 & HD\,24086 & F2 & 6890 & - & - & - & 9.563 & 9.142 & 8.72 & - & 8.214 & 8.045 & 8.009 & 18302720 & 5,6,-,-,-,7,7,8,-,4,4,4 \\
AK\,1B\,560 & HD\,23975 & G0 & 6462 & 4.16 & 0.06 & 10.17 & 10.16 & 9.64 & 9.2 & 8.87 & 8.645 & 8.469 & 8.382 & 18306816 & 5,13,13,13,10,10,10,10,10,4,4,4 \\
AK\,II\,34 & HD\,22444 & G0 & 6030 & - & - & 9.76 & 9.76 & 9.23 & 8.78 & 8.47 & 8.141 & 7.946 & 7.902 & 18305024 & 5,6,-,-,10,10,10,10,10,4,4,4 \\
AK\,II\,359 & TYC\,1798-465-1 & - & 5000\tnd & - & - & - & 11.263 & 10.588 & 10.14 & - & 9.363 & 9.124 & 9.021 & 18304768 & -,-,-,-,-,7,7,8,-,4,4,4 \\
AK\,II\,383 & HIP\,16979 & F8 & 6200 & - & - & - & 10.691 & 10.044 & 9.81 & 9.39\tne & 9.018 & 8.775 & 8.696& 18305536 & 5,6,-,-,-,1,1,8,3,4,4,4 \\
AK\,II\,437 & HD\,22680 & F8 & 6200 & - & - & - & 10.521 & 9.937 & 9.55 & 9.26\tne & 8.92 & 8.686 & 8.634& 18305792 & 6,6,-,-,-,7,7,8,3,4,4,4 \\
AK\,III\,288 & HIP\,16639 & F5 & 6440 & - & - & 10.06 & 10.05 & 9.54 & 9.06 & 8.74 & 8.553 & 8.312 & 8.274& 18305280 & 5,6,-,-,10,10,10,10,10,4,4,4 \\
HII\,25 & HD\,23061 & F5 & 6462 & 4.19 & -0.08 & 9.96& 9.95 & 9.47 & 9.01 & 8.74 & 8.514 & 8.325 & 8.263 & 18311424 & 14,15,16,15,10,10,10,10,10,9,4,4,4\\
HII\,102 & TYC 1799-118-1 & G1 & 6100\tnd & - & - & 11.39 & 11.24 & 10.51 & 9.90 & 9.44 & 9.101 & 8.722 & 8.655& 18311680 & 14,-,-,-,10,10,10,10,10,4,4,4 \\
HII\,1132 & HD\,23514 & G0 & 6545 & 4.21 & -0.02 & 9.95& 9.92 & 9.42 & 8.96 & 8.67 & 8.479 & 8.291 & 8.153& 18303232 & 11,15,13,15,10,10,10,10,10,4,4,4\\
HII\,1139 & HD\,23513 & F5 & 6545 & 4.37 & 0.07 & 9.85& 9.85 & 9.37 & 8.97 & 8.68 & 8.473 & 8.28 & 8.243& 18311936 & 14,17,13,17,10,10,10,10,10,4,4,4\\
HII\,1766 & HD\,23732 & F4 & 6720 & 4.5 & -0.02 & 9.67& 9.60 & 9.13 & 8.70 & 8.41 & 8.137 & 7.912 & 7.862& 18302976 & 14,18,18,18,19,19,19,19,19,4,4,4 \\
HII\,2172 & HD\,282965 & F9 & 6000\tnd & - & - & 11.15 & 11.06 & 10.43 & 9.88 & 9.50 & 9.246 & 9.028 & 8.949& 18312192 & 14,-,-,-,10,10,10,10,10,4,4,4 \\
HII\,3031 & HD\,24132 & F2 & 7000 & 3.98 & 0.09 & 9.3 & 9.25 & 8.87 & 8.49 & 8.27 & 8.059 & 7.93 & 7.877& 18308096 & 14,13,13,13,19,19,19,19,19,4,4,4 \\
PELS\,7 & TYC\,1802-95-1 & - & 5600\tnd & - & - & - & 11.142 & 10.379 & 10.08 & - & 9.24 & 8.953 & 8.833& 18308352 & -,-,-,-,-,7,7,8,-,4,4,4 \\
PELS\,20 & HIP\,17020 & - & 5600\tnd & - & - & - & 11.17 & 10.52 & 10.26 & 9.82\tne & 9.358 & 9.053 & 9.041& 18308608 & -,-,-,-,-,1,1,8,3,4,4,4 \\
PELS\,23 & HIP\,17245 & - & 6000\tnd & - & - & - & 10.673 & 10.102 & 9.72 & 9.44\tne & 8.959 & 8.657 & 8.593& 18308864 & -,-,-,-,-,1,1,8,3,4,4,4 \\
PELS\,25 & HIP\,17125 & F5 & 6440 & - & - & - & 10.088 & 9.571 & 9.45 & 9.00\tne & 8.638 & 8.479 & 8.373 & 18309120 & 5,6,-,-,-,7,7,8,3,4,4,4 \\
PELS\,40 & BD+21 516 & - & 5700\tnd & - & - & - & 10.546 & 9.982 & 9.59 & - & 8.902 & 8.694 & 8.586 & 18307072 & -,-,-,-,-,7,7,2,-,4,4,4 \\
PELS\,86 & HIP\,18544 & F8 & 6200 & - & - & - & 9.928 & 9.377 & 9.04 & 8.81\tne & 8.436 & 8.248 & 8.197 & 18309376 & 5,6,-,-,-,1,1,2,3,4,4,4 \\
PELS\,121 & BD+23 455 & - & 5700\tnd & - & - & - & 10.891 & 10.294 & 10.03 & - & 9.066 & 8.754 & 8.679 & 18309632 & 5,-,-,-,-,1,1,2,-,4,4,4 \\
PELS\,124 & HIP\,16753 & - & 6100\tnd & - & - & - & 10.401 & 9.836 & 9.74 & 9.25\tne & 8.831 & 8.599 & 8.541& 18307328 & -,-,-,-,-,1,1,8,3,4,4,4 \\
PELS\,128 & BD+26 592 & G0 & 6030 & - & - & - & 11.012 & 10.276 & 9.91 & - & 9.003 & 8.778 & 8.66 & 18310144 & 5,6,-,-,-,7,7,8,-,4,4,4 \\
PELS\,135 & TYC\,1256-516-1 & F5 & 6440 & - & - & - & 9.869 & 9.391 & 8.91 & - & 8.404 & 8.225 & 8.152 & 18310400 & 6,6,-,-,-,7,7,8,-,4,4,4 \\
PELS\,146 & HIP\,18091 & - & 5600\tnd & - & - & - & 11.17 & 10.513 & 9.91 & 9.80\tne & 9.267 & 8.981 & 8.873& 18310656 & -,-,-,-,-,20,20,8,3,4,4,4 \\
PELS\,150 & HD\,23935 & F8 & 6200 & - & - & 10.16 & 10.10 & 9.57 & 9.11 & 8.76 & 8.489 & 8.332 & 8.254 & 18310912 & 5,6,-,-,10,10,10,10,10,4,4,4 \\
PELS\,173\tnf & BD+22 617C & - & 6500\tnd & - & - & - & 10.06 & 9.61 & - & - & 8.676 & 8.558 & 8.488 & 18311168 & -,-,-,-,-,21,21,-,-,4,4,4 \\
PELS\,174 & HIP\,18955 & F5 & 6440 & - & - & - & 10.277 & 9.678 & 9.34 & 9.04\tne & 8.519 & 8.262 & 8.183 & 18307840 & 5,6,-,-,-,1,1,8,3,4,4,4 \\
TrS\,42 & HIP\,17316 & G0 & 6030 & - & - & 10.44 & 10.41 & 9.86 & 9.38 & 9.07 & 8.797 & 8.604 & 8.526 & 18306304 & 5,6,-,-,10,10,10,10,10,4,4,4 \\
Tr\,60 & HD\,24302 & F8 & 6450 & 4.21 & -0.2 & 9.91 & 9.93 & 9.45 & 9.01 & 8.72 & 8.517 & 8.303 & 8.259 & 18307584 & 11,22,13,22,10,10,10,10,10,4,4,4 \\
\enddata
\tablenotetext{a}{ Spectral Type}
\tablenotetext{b}{ We estimated log$g$ to be 4.5 when no data were available.}
\tablenotetext{c}{ We estimated [Fe/H] to be 0 when no data were available.}
\tablenotetext{d}{ $T_{eff}$ was estimated using $B-V$ color where derived temperatures from spectra were not available. See references for sources of derived temperatures.}
\tablenotetext{e}{ $I_C$. were estimated from $V-I$ listed in the Hipparcos catalog. They are converted to $I_J$ using the following relation: $(V-I)_C = 0.778(V-I)_J$ \citep{bessell1979}.}
\tablenotetext{f}{ Pels 173: confusion of source. See text for details.}
\tablecomments{The photometry and stellar parameters given in this table served as the primary inputs for the SED fitting, to obtain the predicted flux at 24$\mu$m for each star in the new sample. Several stars also had photometry in other bands: Hipparcos $B$ and $V$ - AK: IA\,36, 1A\,76, IB\,8, II\,383, II\,437, III\,288, III\,756, HII 3031, Pels: 20, 23, 25, 86, 146, 174, TrS42, Tr60; IRAS - HII 1132; uvby - AK\,1B\,560, HII: 25, 1132, 1139, 1766, 3031, Tr60; Johnson $JHK$ - HII 1766, 3031. The other bands were also included in the SED fitting, but are not listed in the table.}
\tablerefs{1 - \citet{kharchenko2004}, 2 - \citet{monet2003}, 3 - \citet{perryman1997}, 4 - \citet{skrutskie2006}, 5 - \citet{roeser1988}, 6 - \citet{wright2003}, 7 - \citet{hog2000}, 8 - \citet{ivanov2008}, 9 - \citet{allende1999}, 10 - \citet{mendoza1967}, 11 - SIMBAD, 12 - \citet{belikov2002}, 13 - \citet{philip1980}, 14 - \citet{soderblom1993}, 15 - \cite{boesgaard1988}, 16 - \citet{gray2001}, 17 - \citet{boesgaard1989}, 18 - \citet{boesgaard1990}, 19 - \citet{iriarte1969}, 20 - \citet{vanleeuwen1986}, 21 - \citet{stauffer2007}, 22 - \citet{suchkov2003}, Hipparcos $BV$ - \citet{perryman1997}, IRAS - \citet{moshir1990}, uvby - \citet{hauck1998}, Johnson $JHK$ - \citet{iriarte1969}}
\end{deluxetable}

\eject

\begin{deluxetable}{lccccccrccc}
\rotate
\tabletypesize{\footnotesize}
\tablecolumns{10} 
\tablewidth{0pt}

\tablecaption{MIPS 24$\mu$m Fluxes and Derived Quantities for the Pleiades}
\tablehead{\colhead{Target}                                          &
           \colhead{$F_\nu$ (error)}                                 &
	   \colhead{Excess Ratio}                                    &
	   \colhead{$\Delta V$}                                      &
	   \colhead{$v$sin$i$}                                       &
	   \colhead{$Log L_X$\tna}                                   &
	   \colhead{$A_*/A_\Sun$\,\tnb}                              &
	   \colhead{$L_{IR}/L_*$}                                    &
	   \colhead{Notes\,\tablenotemark{c}}                        &
	   \colhead{Ref.}                                            \\
	   \colhead{}                                                &
	   \colhead{(mJy)}                                           & 
	   \colhead{}                                                &
	   \colhead{}                                                &
	   \colhead{($km s^{-1}$)}                                   &
	   \colhead{(log($ergs s^{-1}$)}                             &
	   \colhead{}                                                &
	   \colhead{}                                                &
	   \colhead{}                                                &
	   \colhead{}                                                }
\startdata
AK\,IA\,36 & 2.41 (0.02) & 1.03 & 0.42 & 33.9 & - & 1.0 & $<$4.9 $\times 10^{-6}$ & PB & 1,4,-,1 \\
AK\,1A\,56 & 2.53 (0.02) & 1.13 & -0.05 & 14.4 & - & 0.9 & 2.4 $\times 10^{-5}$ & & 1,4,-,- \\
AK\,1A\,76 & 4.59 (0.05) & 1.26 & 0.01 & 11.7 & - & 1.6 & 2.5 $\times 10^{-5}$ & & 1,4,-,- \\
AK\,1A\,317 & 3.17 (0.04) & 1.01 & 0.00 & - & 30.0 & 1.4 & $<$7.9 $\times 10^{-7}$ & & 1,-,7,- \\
AK\,1B\,7 & 3.46 (0.04) & 1.12 & 0.30 & - & - & 1.3 & 1.5 $\times 10^{-5}$ & PB & 1,-,-,1 \\
AK\,IB\,8 & 2.13 (0.02) & 1.04 & -0.10 & 11.9 & - & 1.1 & $<$4.4 $\times 10^{-6}$ & & 1,4,-,- \\
AK\,1B\,146 & 3.62 (0.06) & 0.93 & 0.16 & - & 28.3 & 1.2 & $<$9.2 $\times 10^{-6}$ & & 3,-,7,- \\
AK\,1B\,365 & 3.26 (0.04) & 1.01 & 0.23 & 39.9 & - & 1.3 & $<$7.2 $\times 10^{-7}$ & PB & 1,4,-,1 \\
AK\,1B\,560 & 4.37 (0.04) & 1.34 & 0.11 & 21.6 & - & 1.5 & 3.2 $\times 10^{-5}$ & & 1,4,-,- \\
AK\,1B\,590 & 4.52 (0.05) & 0.99 & -0.01 & - & - & 1.7 & $<$5.9 $\times 10^{-6}$ & & 1,-,-,- \\
AK\,II\,34 & 5.30 (0.05) & 1.06 & 0.58 & 22.1 & 29.9 & 1.4 & $<$1.1 $\times 10^{-5}$ & PB & 1,4,7,12 \\
AK\,II\,359 & 1.82 (0.02) & 0.99 & 0.04 & 16.6 & - & 1.0 & $<$6.5 $\times 10^{-6}$ & & 1,4,-,- \\
AK\,II\,383 & 2.88 (0.02) & 1.17 & 0.39 & 24.5 & 29.7 & 1.3 & 1.5 $\times 10^{-5}$ & PB & 1,4,7,1 \\
AK\,II\,437 & 9.89 (0.09) & 3.79 & 0.18 & 29.6 & - & 1.4 & 2.4 $\times 10^{-4}$ & & 1,4,-,- \\
AK\,III\,288 & 3.57 (0.05) & 0.99 & 0.13 & 36.4 & - & 1.5 & $<$6.3 $\times 10^{-6}$ & & 1,4,-,- \\
HII\,25 & 4.13 (0.05) & 1.12 & -0.02 & 46.0 & 29.1 & 1.6 & 1.1 $\times 10^{-5}$ & & 1,4,7,- \\
HII\,102 & 2.72 (0.02) & 1.04 & 0.44 & 24.58 & - & 0.7 & $<$1.1 $\times 10^{-5}$ & B; sep.= 3.63\arcsec & 1,5,-,11 \\
HII\,120 & 1.69 & 1.04 & -0.04 & 9.4 & 30.2 & 0.8 & $<$6.3 $\times 10^{-6}$ & & 2,4,9,- \\
HII\,152 & 1.82 & 1.12 & 0.02 & 11.3 & 30.6 & 1.0 & 1.3 $\times 10^{-5}$ & & 2,4,9,- \\
HII\,173 & 2.10 & 1.00 & 0.61 & 5.2 & 28.9 & 0.6 & $<$6.9 $\times 10^{-7}$ & SB & 2,4,7,10,11 \\
HII\,174 & 1.32 & 0.99 & -0.06 & 14.4 & 30.1 & 0.6 & $<$1.5 $\times 10^{-5}$ & & 2,4,7,- \\
HII\,250 & 1.90 & 1.12 & -0.06 & 7.1 & 29.4 & 0.9 & 1.4 $\times 10^{-5}$ & & 2,4,7,- \\
HII\,293 & 1.80 (0.05) & 0.98 & 0.01 & 6.6 & 29.2 & 0.8 & $<$9.3 $\times 10^{-6}$ & & 3,4,7,- \\
HII\,314 & 1.98 & 1.02 & -0.01 & 42.8 & 30.3 & 0.9 & $<$2.4 $\times 10^{-6}$ & & 2,4,7,- \\
HII\,405 & 2.94 (0.08) & 0.98 & 0.10 & 18.5 & 29.7 & 1.4 & $<$6.1 $\times 10^{-6}$ & & 3,4,7,- \\
HII\,489 & 2.60 (0.09) & 1.20 & -0.12 & 18.3 & 29.2 & 1.0 & 2.5 $\times 10^{-5}$ & & 3,4,7,- \\
HII\,514 & 2.04 & 1.19 & 0.07 & 10.6 & 28.7 & 0.9 & 2.6 $\times 10^{-5}$ & & 2,4,8,- \\
HII\,571 & 1.67 (0.05) & 1.05 & -0.06 & 7.6 & - & 0.7 & $<$1.1 $\times 10^{-5}$ & & 3,4,-,- \\
HII\,727 & 3.68 (0.08) & 1.00 & 0.29 & 66.4 & 29.8 & 1.2 & $<$4.3 $\times 10^{-7}$ & PB & 3,4,7,12 \\
HII\,739 & 5.09 (0.08) & 1.00 & 0.73 & 14.6 & 30.3 & 0.9 & $<$7.8 $\times 10^{-6}$ & PB & 3,4,7,12 \\
HII\,923 & 2.77 (0.08) & 1.06 & 0.09 & 18.3 & 29.2 & 1.1 & $<$8.3 $\times 10^{-6}$ & & 3,4,7,- \\
HII\,996 & 2.45 (0.07) & 1.21 & -0.04 & 11.9 & 28.9 & 1.1 & 2.1 $\times 10^{-5}$ & & 3,4,7,- \\
HII\,1015 & 1.77 & 0.98 & -0.03 & 9.6 & 29.3 & 1.1 & $<$6.1 $\times 10^{-6}$ & & 2,4,7,- \\
HII\,1101 & 3.30 & 1.48 & 0.00 & 20.0 & 29.6 & 1.1 & 5.3 $\times 10^{-5}$ & & 2,4,7,- \\
HII\,1117 & 3.08 (0.09) & 1.04 & 0.75 & 4.4 & 29.1 & 0.9 & $<$9.8 $\times 10^{-6}$ & SB & 3,4,7,10,11 \\
HII\,1132 & 68.04 (0.36) & 17.44\phn & 0.16 & 49.5 & - & 1.5 & 1.9 $\times 10^{-3}$ & & 1,4,-,- \\
HII\,1139 & 4.03 (0.06) & 1.09\tnd & 0.08 & 31.8 & - & 1.8 & $<$7.4 $\times 10^{-6}$ & & 1,4,-,- \\
HII\,1182 & 1.93 & 1.03 & -0.03 & 16.6 & - & 1.0 & $<$3.2 $\times 10^{-6}$ & & 2,4,-,- \\
HII\,1200 & 3.09 & 1.15 & -0.07 & 13.7 & - & 1.3 & 1.6 $\times 10^{-5}$ & & 2,4,-,- \\
HII\,1309 & 3.55 (0.11) & 0.97 & -0.06 & 85.0 & 29.4 & 1.7 & $<$5.2 $\times 10^{-6}$ & & 3,6,7,- \\
HII\,1338 & 7.34 (0.10) & 0.96 & 0.76 & 110.0 & 29.2 & 1.7 & $<$1.0 $\times 10^{-5}$ & B; sep.= 0.2\arcsec & 3,6,7,13 \\
HII\,1514 & 2.11 (0.07) & 1.07 & 0.06 & 13.7 & 29.2 & 1.0 & $<$8.0 $\times 10^{-6}$ & & 3,4,7,- \\
HII\,1613 & 2.83 (0.08) & 1.01 & 0.01 & 20.0 & 29.3 & 1.4 & $<$1.1 $\times 10^{-6}$ & & 3,4,7,- \\
HII\,1726 & 5.25 (0.10) & 1.04 & 0.66 & 12.9 & 29.5 & 1.4 & $<$6.4 $\times 10^{-6}$ & VB; sep.= 0.5\arcsec & 3,4,7,10 \\
HII\,1766 & 8.45 (0.06) & 1.60 & 0.27 & 23.1 & - & 1.5 & 9.5 $\times 10^{-5}$ & PB & 1,4,-,10 \\
HII\,1797 & 3.68 (0.08) & 1.52 & -0.13 & 19.8 & 29.4 & 1.4 & 4.3 $\times 10^{-5}$ & & 3,4,7,- \\
HII\,1856 & 2.43 (0.07) & 0.94 & -0.01 & 15.6 & 29.3 & 1.3 & $<$5.3 $\times 10^{-6}$ & & 3,4,7,- \\
HII\,1912 & 5.73 (0.12) & 0.97 & 0.35 & 75.0 & 29.5 & 1.5 & $<$1.1 $\times 10^{-5}$ & VB; sep.= 0.3\arcsec & 3,6,7,10 \\
HII\,1924 & 2.24 (0.07) & 1.04 & -0.08 & 14.2 & 29.2 & 1.2 & $<$4.3 $\times 10^{-6}$ & & 3,4,7,- \\
HII\,2027 & 2.24 (0.06) & 0.97 & 0.63 & 2.4 & 29.6 & 0.6 & $<$2.2 $\times 10^{-5}$ & B; sep.:0.10\arcsec & 3,4,7,11 \\
HII\,2147 & 2.62 & 1.00 & 0.53 & 27.1 & 30.4 & 0.6 & $<$1.6 $\times 10^{-6}$ & SB & 2,4,7,11 \\
HII\,2172 & 2.15 (0.02) & 1.10\tnd & -0.08 & - & 29.3 & 1.1 & 9.5 $\times 10^{-6}$ & & 1,-,7,- \\
HII\,2278 & 2.20 & 0.98 & 0.75 & 6.2 & 29.6 & 0.6 & $<$2.2 $\times 10^{-5}$ & B; sep.= 0.27\arcsec & 2,4,7,11 \\
HII\,2506 & 2.06 & 1.00 & -0.07 & 13.9 & 29.6 & 1.1 & $<$3.4 $\times 10^{-8}$ & & 2,4,9,- \\
HII\,2644 & 1.31 & 1.01 & 0.06 & 3.6 & 29.0 & 0.8 & $<$7.4 $\times 10^{-7}$ & & 2,4,7,- \\
HII\,2786 & 2.00 & 0.99 & -0.04 & 22.2 & 29.7 & 1.2 & $<$5.3 $\times 10^{-6}$ & & 2,4,9,- \\
HII\,2881 & 1.68 & 1.00 & 0.69 & 8.0 & 29.9 & 0.5 & $<$1.6 $\times 10^{-7}$ & B; sep.= 0.08\arcsec & 2,4,7,11 \\
HII\,3031 & 5.26 (0.05) & 1.01\tnd & 0.01 & - & 29.5 & 2.0 & $<$1.2 $\times 10^{-6}$ & & 1,-,9,- \\
HII\,3097 & 1.65 & 1.01 & 0.06 & 14.7 & 29.8 & 0.8 & $<$2.6 $\times 10^{-6}$ & & 2,4,9,- \\
HII\,3179 & 2.41 & 1.00 & 0.00 & 5.0 & 29.6 & 1.2 & $<$2.2 $\times 10^{-7}$ & & 2,4,7,- \\
PELS\,7  & 2.16 (0.02) & 0.99 & 0.74 & 2.7 & - & 1.1 & $<$7.3 $\times 10^{-8}$ & PB & 1,4,-,1 \\
PELS\,20 & 2.19 (0.02) & 1.19 & -0.07 & 9.7 & - & 1.1 & 1.7 $\times 10^{-5}$ & & 1,4,-,- \\
PELS\,23 & 2.90 (0.04) & 1.06 & -0.05 & 36.2 & - & 1.1 & $<$8.9 $\times 10^{-6}$ & & 1,4,-,- \\
PELS\,25 & 3.33 (0.04) & 1.02 & 0.15 & - & 29.8 & 1.6 & $<$1.4 $\times 10^{-6}$ & & 1,-,7,- \\
PELS\,40 & 2.77 (0.04) & 1.03 & 0.04 & 11.9 & - & 1.3 & $<$2.8 $\times 10^{-6}$ & & 1,4,-,- \\
PELS\,86 & 4.24 (0.04) & 1.10 & 0.57 & - & - & 1.6 & $<$9.2 $\times 10^{-6}$ & PB & 1,-,-,1 \\
PELS\,121 & 2.45 (0.02) & 0.97 & -0.11 & 4.9 & - & 1.0 & $<$9.7 $\times 10^{-6}$ & & 1,4,-,- \\
PELS\,124 & 3.01 (0.04) & 1.06 & 0.19 & 20.4 & - & 1.4 & $<$5.1 $\times 10^{-6}$ & & 1,4,-,- \\
PELS\,128 & 2.90 (0.10)\tnf & 1.05\tnd & 0.70 & 4.8 & - & 1.0 & $<$8.9 $\times 10^{-6}$ & PB & 1,4,-,1 \\
PELS\,135 & 4.35 (0.04) & 1.10 & 0.05 & - & - & 1.5 & $<$1.1 $\times 10^{-5}$ & & 1,-,-,- \\
PELS\,146 & 3.00 (0.03) & 1.43 & -0.10 & 17.8 & - & 1.0 & 6.3 $\times 10^{-5}$ & & 1,4,-,- \\
PELS\,150 & 4.47 (0.05) & 1.26 & 0.24 & 26.6 & - & 1.4 & 3.0 $\times 10^{-5}$ & PB & 1,4,-,1 \\
PELS\,173 & 3.39 (0.04) & 1.18\tnd & -0.31 & 37.8 & - & 1.8 & $<$1.1 $\times 10^{-5}$ & & 1,4,-,- \\
PELS\,174 & 4.17 (0.04) & 1.06 & 0.52 & 58.9 & 30.0 & 1.1 & $<$1.1 $\times 10^{-5}$ & PB & 1,4,9,1 \\
Tr\,60  & 3.64 (0.05) & 1.01 & 0.00 & - & 29.9 & 1.6 & $<$7.5 $\times 10^{-7}$ & & 1,-,7,- \\
TrS\,42  & 3.17 (0.04) & 1.10 & 0.08 & 19.6 & - & 1.4 & 1.0 $\times 10^{-5}$ & & 1,4,-,- \\
\enddata

\tablenotetext{a}{ Upper limits on $LogL_x$ were not computed due to discrepencies between sources of X-ray photometry.}
\tablenotetext{b}{ where $A_\Sun$ = 6.1 $\times 10^{22} cm^{2}$}
\tablenotetext{c}{ B: binary; PB: photometry binary; SB: spectroscopic binary}
\tablenotetext{d}{ Excess ratios not included in Figure 1. See text for details.}
\tablenotetext{e}{ Aperture photometry}
\tablerefs{In order of: $F_\nu$, $v$sin$i$, $Log L_X$, Notes: 1 - This work, 2 - \citet{stauffer2005}, 3 - \citet{gorlova2006}, 4 - \citet{mermilliod2009}, 5 - \citet{white2007}, 6 - \citet{glebocki2000}, 7 - \citet{flesch2004}, 8 - \citet{marino2008}, 9 - \citet{micela1990}, 10 - \citet{mermilliod1992}, 11 - \citet{bouvier1997}, 12 - \citet{kahler1999}, 13 - \citet{mason2003}}
\tablecomments{The errors reported for the stars in the new sample are random. Systematic errors are approximately 2\% of the measurement. Errors on measurements for stars taken from \citet{stauffer2005,gorlova2006} are estimated to be about the same.}
\end{deluxetable}

\eject

\begin{deluxetable}{lcccc}
\tabletypesize{\footnotesize}
\tablecolumns{5}
\tablewidth{0pt}
\tablecaption{Fit Parameters for Isochrones}
\tablehead{\colhead{Cluster}     &
           \colhead{Age}         &
	   \colhead{[Fe/H]}      &
	   \colhead{$A_V$}       &
	   \colhead{References}  }

\startdata
Pleiades & 115 & +0.03 & 0.12 & 1,2,3 \\
NGC 2547 & 35 & -0.26 & 0.19 & 4,5,5 \\
Blanco 1 & 90 & +0.04 & 0.06 & 6,7,8 \\
\enddata
\tablerefs{1 - \citet{meynet1993,stauffer1998,martin2001}, 2 - \citet{soderblom2009}, 3 - \citet{stauffer1987}, 4 - \citet{jeffries2005}, 5 - \citet{claria1982}, 6 - \citet{panagi1997}, 7 - \citet{ford2005}, 8 - \citet{westerlund1988}}
\end{deluxetable}

\eject

\begin{deluxetable}{llccccccr}
\rotate
\tabletypesize{\footnotesize}
\tablecolumns{8}
\tablewidth{0pt}
\tablecaption{MIPS 24$\mu$m Magnitudes, Excess Ratios, Mass Loss Rates and Fractional Luminosities for NGC 2547}
\tablehead{\colhead{RA}                                        &
           \colhead{DEC}                                       &
           \colhead{$[24]$}                                    &
           \colhead{$V-K_S$}                                   &
	   \colhead{Excess Ratio}                              &
	   \colhead{$Log L_X$}                                 &
	   \colhead{$A_*/A_\Sun$}                              &
	   \colhead{$L_{IR}/L_*$}                              \\
	   \colhead{}                                          &
	   \colhead{}                                          &
	   \colhead{(mag)}                                     &
	   \colhead{(mag)}                                     &
	   \colhead{}                                          &
	   \colhead{(Log($ergs s^{-1}$)}                       &
	   \colhead{}                                          &
	   \colhead{}                                          }
\startdata
122.345375 & -49.13269444 & 10.26 & 1.31 & 1.78 & 29.4 & 1.1 & 8.9 $\times 10^{-5}$ \\
122.446875 & -49.21808333 & 10.67 & 1.48 & 1.15 & 30.4 & 0.9 & 2.7 $\times 10^{-5}$ \\
122.47525 & -49.32247222 & 10.97 & 1.79 & 0.94 & 29.3 & 0.7 & $<$1.7 $\times 10{-5}$ \\  
122.5161667 & -49.01861111 & 10.85 & 1.65 & 1.04 & 30.4 & 0.8 & $<$9.2 $\times 10^{-6}$ \\
122.538704 & -49.348375 & 10.74 & 1.64 & 1.07 & $<$29.3 & 0.8 & $<$1.8 $\times 10^{-5}$ \\
122.539 & -49.01541667 & 10.35 & 2.06 & 1.08 & 30.7 & 0.6 & $<$6.2 $\times 10^{-5}$ \\
122.5485417 & -49.37413889 & 10.22 & 1.41 & 0.97 & 29.1 & 1.0 & $<$6.2 $\times 10^{-5}$ \\  
122.556 & -49.34547222 & \phn9.65 & 1.26 & 1.05 & 29.0 & 1.2 & $<$1.5 $\times 10^{-5}$ \\
122.5642083 & -49.09686111 & 10.54 & 1.88 & 1.12 & 30.1 & 0.6 & 5.5 $\times 10^{-5}$ \\
122.576542 & -49.112867 & 10.01 & 1.43 & 1.51 & $<$29.3 & 1.0 & 1.1 $\times 10^{-4}$ \\
122.605208 & -49.316325 & 10.72 & 1.45 & 0.88 & $<$29.3 & 1.0 & $<$1.1 $\times 10^{-5}$ \\  
122.612625 & -49.14708333 & 11.29 & 1.84 & 1.06 & 30.1 & 0.7 & $<$1.3 $\times 10^{-5}$ \\
122.615 & -49.20263889 & \phn9.81 & 1.31 & 0.90 & 28.7 & 1.1 & $<$1.8 $\times 10^{-5}$ \\  
122.6446667 & -49.14447222 & 10.85 & 1.47 & 1.07 & 30.2 & 1.0 & $<$1.1 $\times 10^{-5}$ \\
122.665958 & -49.077206 & 10.29 & 1.30 & 1.01 & $<$29.3 & 1.2 & $<$1.6 $\times 10^{-6}$ \\
122.6906667 & -49.01888889 & 10.28 & 1.40 & 1.51 & 29.7 & 1.0 & 8.3 $\times 10^{-5}$ \\
122.7079167 & -49.19052778 & 10.28 & 1.33 & 1.16 & 29.7 & 1.4 & 1.9 $\times 10^{-5}$ \\
122.750438 & -49.112367 & 10.65 & 1.76 & 2.21 & $<$29.3 & 0.7 & 2.0 $\times 10^{-4}$ \\
122.758417 & -49.269381 & 10.90 & 1.81 & 1.45 & $<$29.3 & 0.7 & 1.0 $\times 10^{-4}$ \\
122.7596667 & -49.27327778 & 10.47 & 1.81 & 1.08 & 30.4 & 0.7 & $<$3.4 $\times 10^{-5}$ \\
\enddata
\end{deluxetable}

\eject

\begin{deluxetable}{lcccc}
\tabletypesize{\footnotesize}
\tablecolumns{4}
\tablewidth{0pt}
\tablecaption{Derived Quantities for Blanco 1}
\tablehead{\colhead{Name}                        &
           \colhead{$A_*/A_\Sun$}                &
           \colhead{Log $L_X$}                   &
           \colhead{$L_{IR}/L_*$}                }
\startdata
W\,38 & 1.6 & 29.85 & 1.5 $\times 10^{-5}$ \\
W\,53 & 1.2 & 29.58 & 2.0 $\times 10^{-5}$ \\
W\,56 & 1.3 & 29.4 & $<$1.1 $\times 10^{-7}$ \\
W\,58 & 1.5 & 29.74 & $<$1.2 $\times 10^{-6}$ \\
W\,68 & 1.5 & 28.95 & $<$6.4 $\times 10^{-6}$ \\
W\,70 & 1.2 & - & $<$8.3 $\times 10^{-6}$ \\
W\,79 & 1.0 & - $<$29.59 & $<$9.8 $\times 10^{-6}$ \\
W\,89 & 1.1 & - & $<$7.7 $\times 10^{-6}$ \\
W\,91 & 1.1 & 29.38 & 4.5 $\times 10^{-5}$ \\
W\,99 & 1.4 & $<$29.3 & 4.3 $\times 10^{-5}$ \\
W\,113 & 1.1 & 29.31 & $<$1.1 $\times 10^{-5}$ \\
ZS\,62 & 0.7 & 29.37 & $<$2.2 $\times 10^{-6}$ \\
ZS\,83 & 0.6 & 29.4 & $<$1.9 $\times 10^{-5}$ \\
ZS\,95 & 0.6 & 28.96 & $<$3.8 $\times 10^{-7}$ \\
ZS\,100 & 0.6 & - & $<$1.8 $\times 10^{-5}$ \\
ZS\,102 & 0.8 & - & 2.7 $\times 10^{-5}$ \\
ZS\,161 & 0.7 & $<$29.43 & $<$1.7 $\times 10^{-7}$ \\
M\,348 & 0.7 & - & $<$1.7 $\times 10^{-5}$ \\
\enddata
\tablerefs{Photometry used is from \citet{stauffer2009} and references therein.}
\end{deluxetable}

\eject

\begin{figure*}
\figurenum{1}
\label{fig1}
\plotone{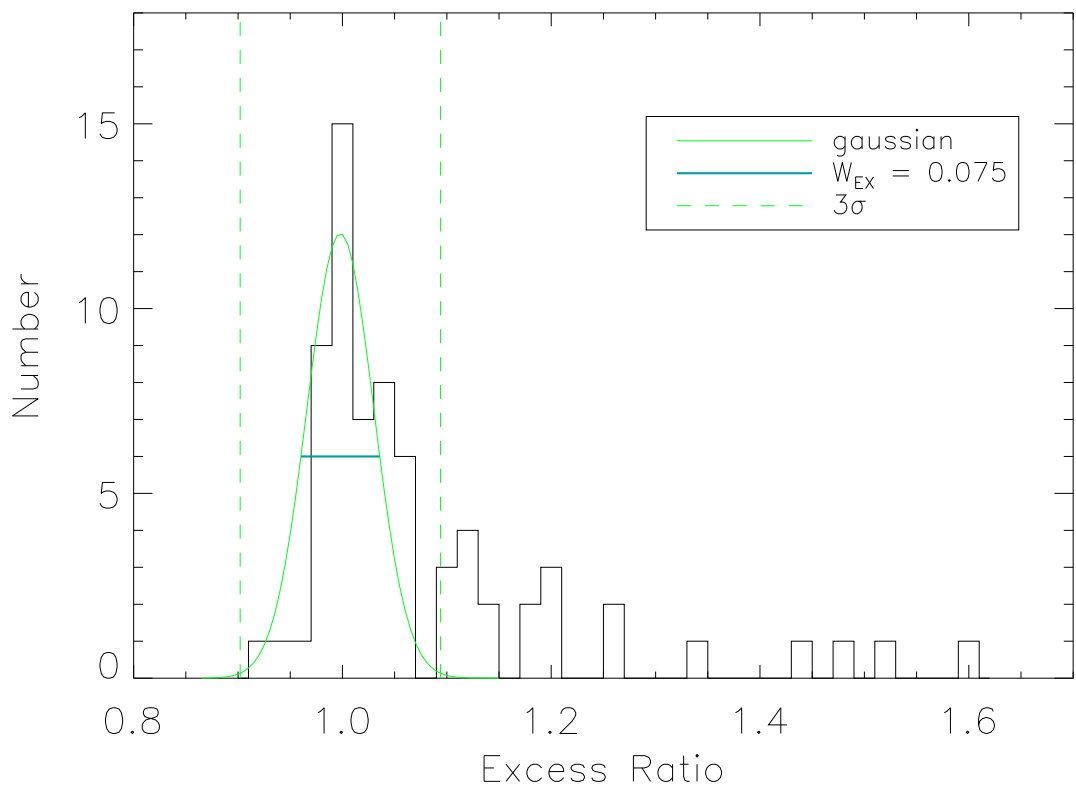}
\caption{Distribution of 24 $\mu$m excess ratios for Pleiades members in Table 2. A Gaussian has been fitted to show the photospheric behavior along with the effect of the errors in extrapolating it to 24 $\mu$m, with the width of the peak, $W_{EX}$, indicated in the plot. The stars with values above and to the right of the Gaussian show infrared excess emission. Two stars with extreme excesses, AKII437 and HII 1132, are not included in the figure.}
\end{figure*}

\eject

\begin{figure*}
\figurenum{2}
\label{fig2}
\plotone{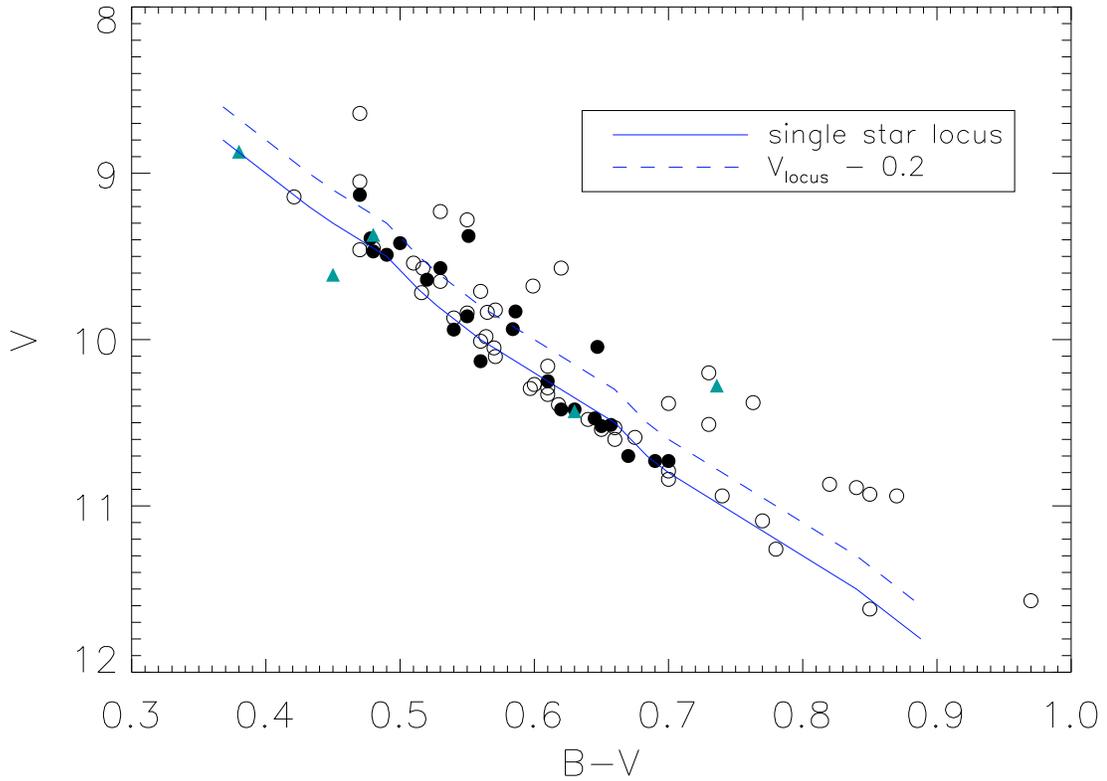}
\caption{$V$ vs. $B-V$ for the Pleiades combined sample, listed in Table 2. The triangles represent HII 1139, HII 2172, HII 3031, Pels 128, and Pels 173, which were not included in the final analysis (see text). The circles represent the rest of the stars, with the filled ones having 24 $\mu$m excesses  $>$ 3$\sigma$. The lines represent the main sequence locus for the Pleiades (solid), and the cutoff for binarity (dashed).}
\end{figure*}

\eject

\begin{figure*}
\figurenum{3}
\label{fig3}
\plotone{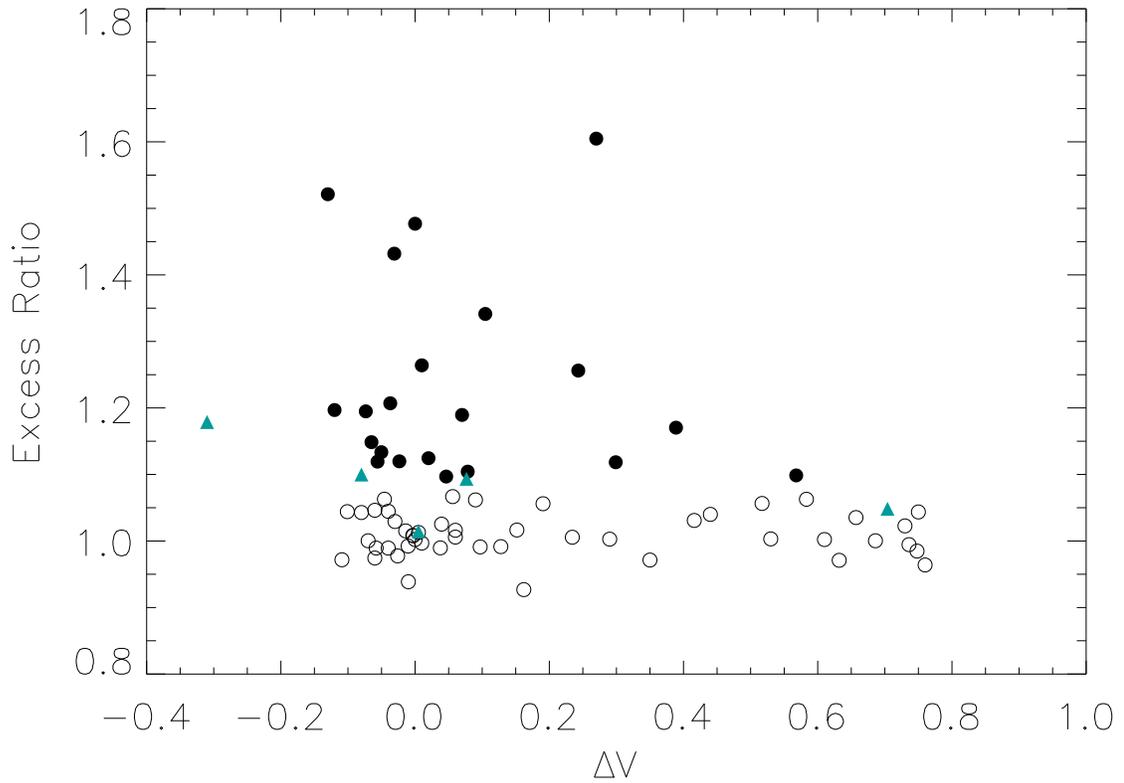}
\caption{Excess Ratio vs. $\Delta V$ for the Pleiades combined sample, where $\Delta V$ is the difference between the observed V magnitude and the main sequence. There is an anticorrelation between 24 $\mu$m excess and binarity. The triangles represent the five stars removed from analysis (HII 1139, HII 2172, HII 3031, Pels 128, and Pels 173). AKII437 and HII 1132 are not shown.}
\end{figure*}

\eject

\begin{figure*}
\figurenum{4}
\label{fig4}
\plotone{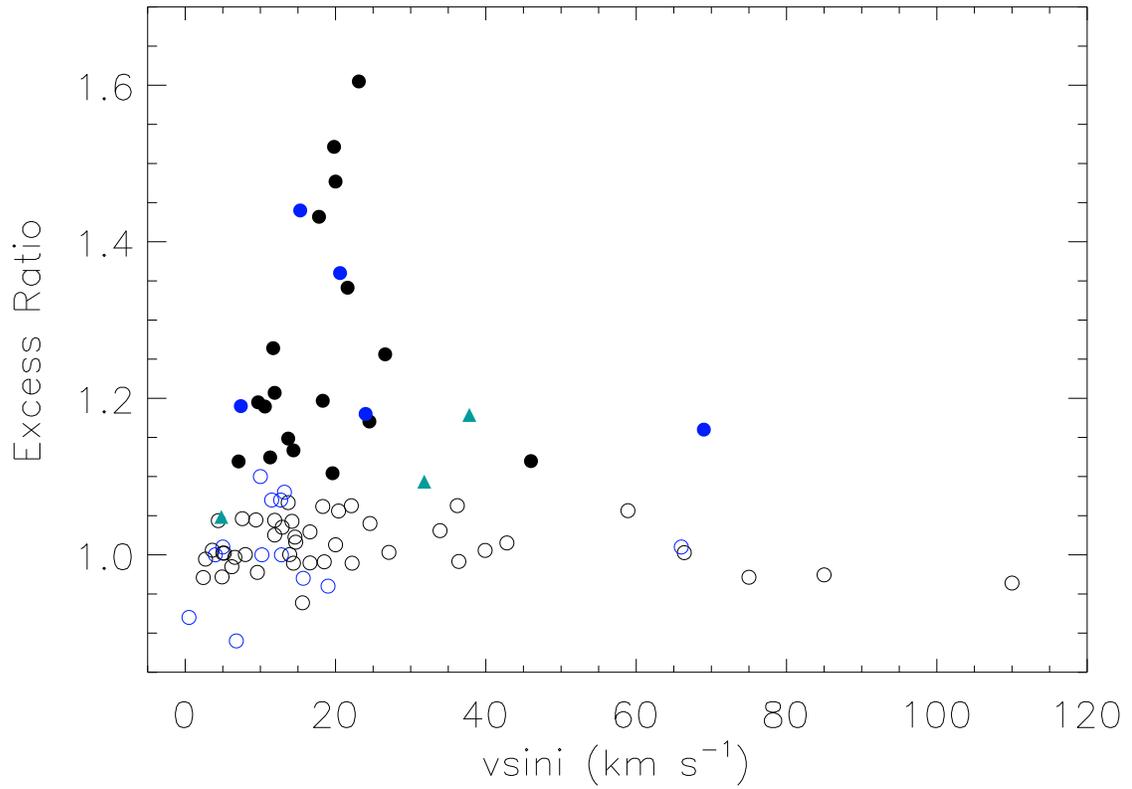}
\caption{Excess Ratio vs. $v$sin$i$ for the Pleiades and Blanco 1. Pleiades stars are in black; Blanco 1 stars in blue. Stars removed from the Pleiades sample (HII 1139, Pels 128, and Pels 173) are represented with triangles. AKII437 and HII 1132 are not shown.}
\end{figure*}

\eject

\begin{figure*}
\figurenum{5}
\label{fig5}
\plotone{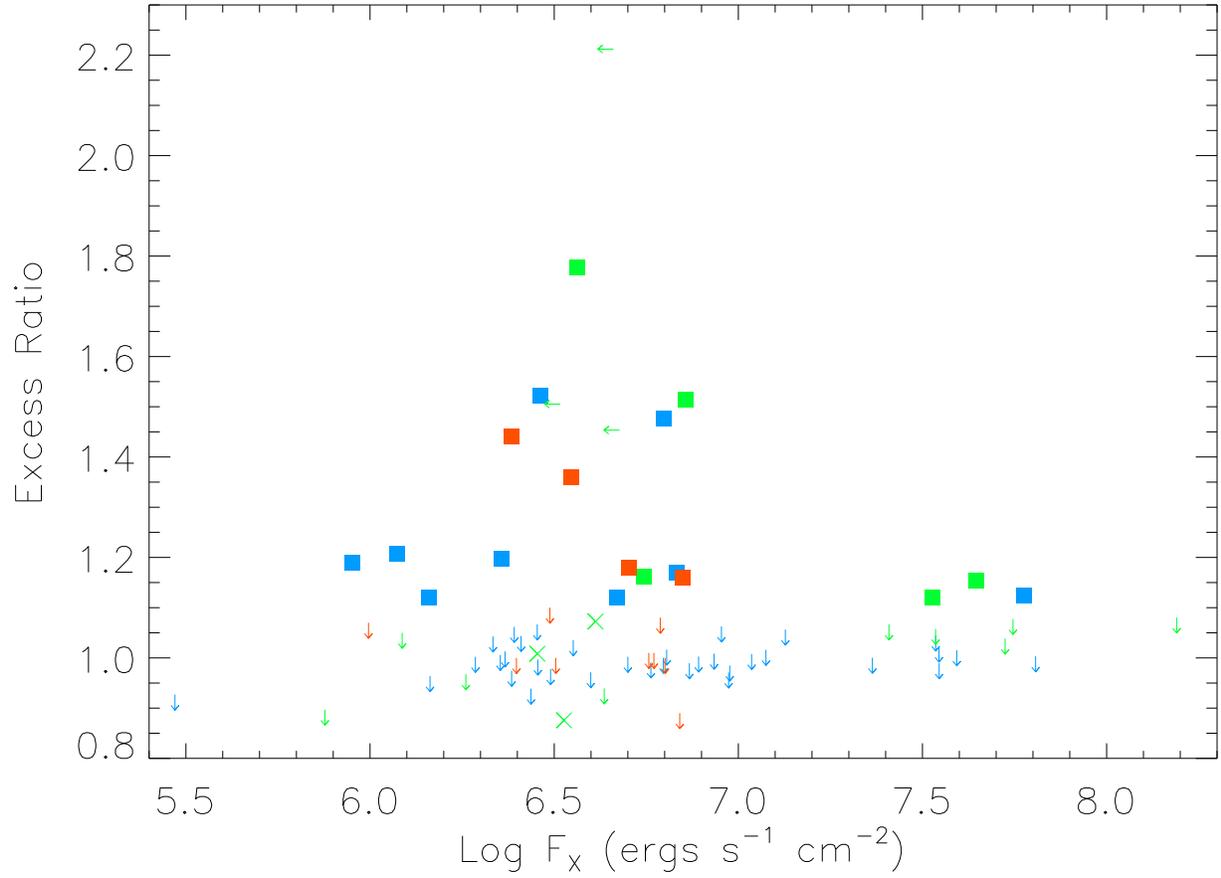}
\caption{Excess Ratio vs. Log X-ray surface brightness (Log $F_X$) for the Pleiades, NGC 2547, and Blanco 1 clusters. Stars with excesses are represented by filled-in squares. Upper limits on $F_X$ are represented by left-pointing arrows; upper limits on excess ratios by down-pointing arrows. Upper limits on both quantities are shown with cross-signs. Pleiades stars are shown in blue, NGC 2547 in green, and Blanco 1 in red.}
\end{figure*}

\end{document}